
  \def\eth{\hbox{$\partial$\kern-0.25em\raise0.6ex\hbox{\rm\char'40}}}

\font\lbf=cmbx10 scaled\magstep2

\def\bs{\bigskip}
\def\ms{\medskip}
\def\np{\vfill\eject}

\def\ni{\noindent}
\def\cl{\centerline}

\def\title#1{\cl{\lbf #1}}
\def\ref#1#2#3#4{#1\ {\it#2\ }{\bf#3\ }#4\par}
\def\refb#1#2#3{#1\ {\it#2\ }#3\par}
\def\ANY{Ann.\ N.Y.\ Acad.\ Sci.}
\def\CQG{Class.\ Qu.\ Grav.}

\def\JMP{J.\ Math.\ Phys.}
\def\PR{Phys.\ Rev.}

\def\PRS{Proc.\ Roy.\ Soc.\ Lond.}

\def\mod#1{\left\vert#1\right\vert^2}
\def\norm#1{\left\vert\left\vert#1\right\vert\right\vert^2}
\def\O#1{\left.#1\right\vert_S}
\def\up#1{{#1\over{\chi\bar\chi}}}
\def\A{\sqrt{{A\over{16\pi}}}}
\def\B{{A^{1/2}\over{(4\pi)^{3/2}}}}
\def\l{\lim_{r\to\infty}}
\def\ll{\lim_{R\to\infty}}
\def\s{\hat{*}}
\def\I{\int_S{*}}
\def\II{\int{\s}}
\def\E{E_\infty}
\def\thorn{I\kern-0.4em\raise0.35ex\hbox{\it o}}
\def\e{\hbox{e}}
\def\D{{\cal D}}
\def\L{{\cal L}}
\def\R{{\cal R}}
\def\scri{{\cal I}}
\def\half{{\textstyle{1\over2}}}
\def\quart{{\textstyle{1\over4}}}

\magnification=\magstep1

\title{Quasi-localisation of Bondi-Sachs energy loss}
\bs\cl{\bf Sean A. Hayward}
\ms\cl{Department of Physics}
\cl{Kyoto University}
\cl{Kyoto 606-01}
\cl{Japan}
\bs\cl{27th May 1994}
\bs\ni
{\bf Abstract.}
A formula is given for the variation of the Hawking energy
along any one-parameter foliation of compact spatial 2-surfaces.
A surface for which one null expansion is positive and the other negative
has a preferred orientation,
with a spatial or null normal direction being called outgoing or ingoing
as the area increases or decreases respectively.
A natural way to propagate such a surface through a hypersurface
is to choose the foliation
such that the null expansions are constant over each surface.
For such uniformly expanding foliations,
the Hawking energy is non-decreasing in any outgoing direction,
and non-increasing in any ingoing direction,
assuming the dominant energy condition.
It follows that the Hawking energy is non-negative
if the foliation is bounded at the inward end
by either a point or a marginal surface,
and in the latter case satisfies the Penrose-Gibbons isoperimetric inequality.
The Bondi-Sachs energy may be expressed as
a limit of the Hawking energy at conformal infinity,
and the energy-variation formula reduces at conformal infinity
to the Bondi-Sachs energy-loss formula.
\bs\cl{PACS: 02.40.Ft, 04.70.Bw, 04.20.Ha, 04.20.Dw}
\np\ni
{\bf I. Introduction}
\ms\ni
One of the classic successes in General Relativity was the realisation that
the total energy of an asymptotically flat space-time
could be measured at conformal infinity,
and that this Bondi-Sachs energy is non-decreasing to the future [1--2].
The energy loss is interpreted as being due to
gravitational radiation escaping to infinity,
so that one may conclude that gravitational radiation carries positive energy.
However, this is restricted to the idealised context
of an asymptotically flat space-time,
with the energy being measured at infinity.
So it is of interest to localise the definition of energy
and the corresponding energy loss.

It is widely accepted that General Relativity does not admit
a purely local gravitational energy,
defined in terms of the geometry in a neighbourhood of a point.
Instead, Penrose has suggested the possibility of
a {\it quasi-local} gravitational energy,
referring to a compact (usually spherical) spatial 2-surface [3].
This is motivated by the character of the Bondi-Sachs energy,
which is defined on spherical spatial 2-surfaces at conformal infinity.
Such a quasi-local energy should reduce to the Bondi-Sachs energy
at conformal infinity,
and should have a monotonicity property
which generalises the Bondi-Sachs energy loss.
Physically, one would expect gravitational radiation to be meaningful
and to carry positive energy
at least in a `wave zone' neighbouring conformal infinity,
and hopefully in a much wider regime.

This article establishes a monotonicity property of the Hawking energy [4]
for a generic class of surfaces.
A positivity property follows directly,
establishing the Penrose-Gibbons isoperimetric inequality
for the Hawking energy.
The Bondi-Sachs energy may be written as a limit of the Hawking energy,
and the energy-loss formula for the Hawking energy
reduces at conformal infinity to the Bondi-Sachs energy-loss formula.
This also provides a particularly straightforward derivation of
the Bondi-Sachs energy loss.

The type of surface to be considered is introduced in Section II.
The energy-variation formula is derived in Section III,
where the monotonicity and positivity theorems are given.
The formula is re-derived in Section IV in terms of spin coefficients,
which enables the limit at conformal infinity to be taken in Section V
using the standard spin-coefficient description.
An alternative way of deriving the Bondi-Sachs energy loss
from the Hawking energy loss is described in Section VI.
Some questions of interpretation are discussed in the Conclusion.
\bs\ni
{\bf II. Mean convex surfaces}
\ms\ni
Quasi-local energy is associated with
compact spatial 2-surfaces embedded in space-time,
and its variation with foliations of such surfaces.
Since a spatial 2-surface has two preferred normal directions,
the null directions,
it is natural to describe the geometry in terms of
two intersecting foliations of null 3-surfaces.
This double-null formalism has been described in detail previously [5--6],
and is summarised as follows.

Denoting the space-time metric by $g$,
and labelling the null 3-surfaces by $\xi_\pm$, one has:
the evolution vectors $u_\pm=\partial/\partial\xi_\pm$,
the normal 1-forms $n_\pm=-\hbox{d}\xi_\pm$,
which are null, $g^{-1}(n_\pm,n_\pm)=0$;
the normalisation $\e^f=-g^{-1}(n_+,n_-)$,
2-metric $h=g+2\e^{-f}n_+\otimes n_-$
and shift 2-vectors $r_\pm=\bot(u_\pm)$,
where $\bot$ indicates projection by $h$;
the Ricci scalar $\R$, covariant derivative $\D$ and Hodge operator $*$ of $h$;
and the expansions $\theta_\pm=*\L_\pm{*}1$,
shears $\sigma_\pm=\bot(\L_\pm h)-\theta_\pm h$,
inaffinities $\nu_\pm=\L_\pm f$
and twist $\omega=\half\e^fh([l_-,l_+])$,
where $\L_\pm$ denotes
the Lie derivative along the null normal vectors $l_\pm=u_\pm-r_\pm$.
The space-time is assumed time-orientable,
and $l_\pm$ are assumed future-pointing.
Additionally, with respect to the spatial 2-metric $h$,
introduce the scalar product of 1-forms,
$\alpha\cdot\beta=h^{-1}(\alpha,\beta)$,
the norm of 1-forms, $\mod\alpha=\alpha\cdot\alpha$,
and the norm of bilinear forms, $\norm\sigma$.

A monotonicity statement requires the surface to have a preferred orientation.
That is, if one wishes to say that some quantity is increasing outwards,
one needs to distinguish between ingoing and outgoing directions.
There is no such preferred orientation for arbitrary orientable surfaces,
so that some restriction on the class of surfaces is required.
For instance,
one would expect the area to increase outwards and decrease inwards.
Applying this locally to the area form $*1$,
one wants the outgoing expansion to be positive,
and the ingoing expansion to be negative.
\ms\ni
{\it Definition.}
A {\it mean convex surface}\footnote\dag
{The name comes from the theory of surfaces in Euclidean 3-space,
where there is a single expansion $\theta$ and shear $\sigma$,
with principal curvatures $\kappa_\pm=\theta\pm||\sigma||/\sqrt2$,
Gaussian curvature $\kappa_+\kappa_-=\theta^2-\half\norm\sigma=\half\R$
and mean curvature $\half(\kappa_++\kappa_-)=\theta$.
A {\it convex} surface $S$ is usually defined by $\O\R>0$,
so that both principal curvatures are positive, fixing an orientation.
Requiring only that the mean curvature be positive
leads to the definition of mean convex surface $S$, $\O{\theta^2}>0$.
Clearly mean convexity implies convexity in this context.
Mean convex surfaces have also been described as {\it well oriented} [7].}
is a compact orientable spatial 2-surface $S$
on which $\O{\theta_+\theta_-}<0$.
Taking $\O{\theta_+}>0$ henceforth,
$l_+$ is called the {\it outgoing} null normal vector,
and $l_-$ the {\it ingoing} null normal vector.
Similarly, any spatial vector $z$ normal to $S$
is called {\it outgoing} if $g(z,l_+)>0$,
or equivalently $g(z,l_-)<0$,
and is called {\it ingoing} if $g(z,l_-)>0$,
or equivalently $g(z,l_+)<0$.
\ms
Mean convex surfaces are generic in the sense that
the set of mean convex surfaces is open
in the space of embedded spatial 2-surfaces.
This space is given by (the sum over compact orientable topologies of $S$ of)
the fibre bundle over $S$
with sections $(h,\theta_\pm,\sigma_\pm,\omega)$,
modulo diffeomorphisms of $S$ and rescalings of the null normals,
$l_\pm\mapsto\e^{\lambda_\pm}l_\pm$.
A point in this space with $\O{\theta_+\theta_-}<0$
is surrounded by a neighbourhood with the same property.
Similarly, any surface sufficiently close to a mean convex surface
in the space-time is also mean convex.
Also,
any point in any space-time is surrounded by neighbouring mean convex surfaces,
since space-time looks sufficiently flat for sufficiently small surfaces.

It should be remarked that
while the signs of $\theta_\pm$ are invariantly defined,
their actual values are not,
since $\theta_\pm\mapsto\e^{\lambda_\pm}\theta_\pm$
under the above rescaling.
This freedom can therefore be used to normalise $\O{\theta_\pm}$
to constants on a mean convex surface $S$, as will be done shortly.

Mean convex surfaces are in some sense an opposite of trapped surfaces [8],
for which $\O{\theta_+\theta_-}>0$.
For instance, in spherical symmetry the expansions are constant over a sphere,
and so any metric sphere is either trapped, marginal or mean convex.
More generally, in an asymptotically flat space-time,
any surface sufficiently close to conformal infinity $\scri$
is mean convex,
because $r^2\theta_+\theta_-|_\scri<0$ for any cut of $\scri$,
where $r$ asymptotes to the area radius.
More generally still,
any surface outside and sufficiently close to a compact outer marginal surface
is also mean convex [6],
so that such surfaces can always be found near a black or white hole.
So it seems likely that mean convex surfaces can be used
to foliate a wide class of space-times outside their trapped regions.

Consider a particular spatial or null hypersurface
containing a mean convex surface.
There are many ways to propagate the surface through the hypersurface,
giving different foliations.
Some of these will not propagate the convexity of the surface.
For instance, a metric sphere in flat space-time can be propagated inwards
to give a non-convex surface of larger area.
The freedom to propagate the surface is given by
rescalings of the null normals,
and as explained above,
there is a natural way to fix this freedom for a mean convex surface $S$,
which is to choose the magnitude of the null normals so that
$\O{\theta_+}$ and $\O{\theta_-}$ are constants.
One could also fix the constants, but this is irrelevant to the applications.
Noting that $\e^f\theta_+\theta_-$ is invariant under such rescalings,
this procedure fixes $\O f$ up to an additive constant.
\ms\ni
{\it Definition.}
A foliation of spatial 2-surfaces $\{S\}$ is described as
{\it uniformly expanding} if
$\O{\D\theta_+}=\O{\D\theta_-}=0$,
where the double-null foliation is adapted to the 2-surfaces.
The foliation is described as a uniformly expanding variation of a given $S$.
\ms
It should be emphasised that
this does not restrict mean convex surfaces further,
since any mean convex surface admits a uniformly expanding variation
in any chosen direction.
Rather, this selects a preferred propagation of the surface
in the chosen direction.
Uniformly expanding foliations have various nice properties.
If such a foliation crosses a trapping horizon [6],
it necessarily does so across a marginal surface.
So this provides a natural way to develop the space-time
from a trapping horizon,
and allows the methods used below to be extended to marginal surfaces.
A uniformly expanding foliation can also be developed from a regular centre.
In an asymptotically flat space-time,
uniformly expanding foliations can be used to develop from $\scri$.
In a rough sense, varying a mean convex surface by uniform expansion
tends to smooth out inhomogeneities in the surface,
at least in something close to flat space-time.
These properties suggest the following conjecture,
whose relevance for the cosmic censorship hypothesis
is discussed in the Conclusion.
\ms\ni
{\it Convexity conjecture.}
If a spatial hypersurface extends to $\scri$ or $i^0$,
does not cross a trapped surface,
and is either $S^2\times R$ with a marginal surface as inner boundary,
or $R^3$, then the hypersurface is coverable by 
a uniformly expanding foliation of mean convex surfaces.
\bs\ni
{\bf III. Variation of the Hawking energy}
\ms\ni
The Hawking energy $E$ of a compact orientable spatial 2-surface $S$
is defined by [4]:
$$8\pi E=\A\I(\R+\e^f\theta_+\theta_-)\eqno(1)$$
with units $G=1$,
where the area is
$$A=\I1.\eqno(2)$$
The Hawking energy is an essentially quasi-local quantity,
in that it cannot be expressed as the integral of a local energy density,
due to the overall factor involving the area.

The variation along a foliation is given by the Lie derivative $\L_z$
along a vector $z$ tangent to the hypersurface and normal to the surfaces.
Since $z$ is a linear combination of the null normals,
$z=\beta l_+-\alpha l_-$,
and $\L_zE=\beta\L_+E-\alpha\L_-E$,
it suffices to calculate $\L_\pm E$.
Using the Gauss-Bonnet theorem
$$\I\R=8\pi(1-\gamma)\eqno(3)$$
where $\gamma$ is the genus or number of handles of $S$,
one has
$$\L_\pm\I\R=0\eqno(4)$$
since there always exists a neighbourhood in which
the topology does not change.
Also,
$$\L_\pm A=\I\theta_\pm.\eqno(5)$$
Of course, $\L_+A>0$ and $\L_-A<0$ for a mean convex surface.
Varying the other terms in $E$ requires
certain of the Einstein equations,
namely the focussing equations [6]:
$$\eqalignno
{&\L_\pm\theta_\pm+\nu_\pm\theta_\pm+\half\theta_\pm^2+\quart\norm{\sigma_\pm}
=-8\pi\phi_\pm&(6a)\cr
&\L_\pm\theta_\mp+\theta_+\theta_-+\e^{-f}\left(
\half\R-\mod{\half\D f\pm\omega}+\D\cdot(\half\D f\pm\omega)\right)
=8\pi\rho&(6b)\cr}$$
where $\phi_\pm=T(l_\pm,l_\pm)$
and $\rho=T(l_+,l_-)$
in terms of the energy tensor $T$.
The conditions
$$\eqalignno
{&\phi_\pm\ge0&(7a)\cr
&\rho\ge0&(7b)\cr}$$
follow from the dominant energy condition [9].
A straightforward calculation yields
$$\eqalignno
{8\pi\L_\pm E&=\A\left[
{1\over{2A}}\I\theta_\pm\I\left(\R+\e^f\theta_+\theta_-\right)
-\I\e^f\theta_\mp\left(\quart\norm{\sigma_\pm}+8\pi\phi_\pm\right)\right.&\cr
&\qquad\left.-\I\theta_\pm\left(\half\R+\half\e^f\theta_+\theta_-
-\mod{\half\D f\pm\omega}+\D\cdot(\half\D f\pm\omega)-8\pi\e^f\rho\right)
\right].&(8)\cr}
$$
\ms\ni
{\it Monotonicity theorem I.}
(i) For a uniformly expanding variation of a surface $S$,
the Hawking energy varies according to
$$8\pi\L_\pm E=
\A\I\left[\theta_\pm\left(\mod{\half\D f\pm\omega}+8\pi\e^f\rho\right)
-\e^f\theta_\mp\left(\quart\norm{\sigma_\pm}+8\pi\phi_\pm\right)\right].
\eqno(9)
$$
(ii) For a mean convex surface $S$,
the dominant energy condition implies that
$E$ is non-decreasing in the outgoing null direction, $\L_+E\ge0$,
is non-increasing in the ingoing null direction, $\L_-E\le0$,
and so is non-decreasing in any outgoing spatial direction $z$, $\L_zE\ge0$,
and non-increasing in any ingoing spatial direction.
\ms\ni
{\it Proof.}
(i) Set $\O{\theta_\pm}$ to constants in (8),
and use the Gauss divergence theorem, $\I\D\cdot\alpha=0$.
(ii) Use the signs $\O{\theta_+}>0$, $\O{\theta_-}<0$,
$\phi_\pm\ge0$ and $\rho\ge0$ (7).
\ms
The result uses a cancellation that occurs for uniformly expanding variations,
and does not generalise to arbitrary variations.
The second bracketed term in (9) corresponds to
the integrand occurring in the Bondi-Sachs energy-loss formula,
as will be shown subsequently,
while the first bracketed term disappears in the limit.
Note that the variation of the Hawking energy is essentially quasi-local,
in that it cannot be expressed as the integral of a local energy flux,
due to the overall factor involving the area.
\ms\ni
{\it Positivity theorem.}
For a uniformly expanding foliation of mean convex surfaces
bounded at the inward end by
either a marginal surface with area $A_0$,
or a point, $A_0=0$,
the Hawking energy of each surface is bounded below by
$$E\ge\sqrt{A_0/16\pi}\ge0\eqno(10)$$
assuming the dominant energy condition.
\ms\ni
{\it Proof.}
Recall that a marginal surface is defined by
$\O{\theta_+}$ or $\O{\theta_-}$ vanishing.
A marginal surface which bounds the inward end
of a foliation of mean convex surfaces is necessarily of the outer type [6],
and being compact, is necessarily spherical [6], $\gamma=0$.
Thus its Hawking energy is $E_0=\sqrt{A_0/16\pi}$,
using the Gauss-Bonnet theorem (3).
Correspondingly, $E_0=0$ for a point,
since $E\to0$ as a surface is shrunk to a point [10].
Then the monotonicity theorem shows that $E\ge E_0$ along the foliation.
\ms
Moreover, one could write $E$ explicitly as $E_0$ plus an integral, using (9).
Note also that
the Hawking energy is manifestly positive for a trapped surface $S$,
for which $\O{\theta_+\theta_-}>0$ and $\gamma=0$.
Thus the theorem provides a positivity property of the Hawking energy
in a neighbourhood of either a point or a black or white hole,
defined in the sense of an outer trapping horizon [6].
Of course, the neighbourhood may not cover the whole space-time,
since the foliation may break down somewhere.
The convexity conjecture would imply that
such a breakdown can be avoided in an asymptotically flat space-time.

The inequality (10) is the Penrose-Gibbons isoperimetric inequality [11--12]
for the Hawking energy.
It is conjectured to hold for the Arnowitt-Deser-Misner energy [11--12]
and the Bondi-Sachs energy [13].
Again, this would follow from the convexity conjecture,
as discussed in the Conclusion.
\bs\ni
{\bf IV. Spin-coefficient form}
\ms\ni
The theory of conformal infinity
was initially constructed using coordinate methods [1--2],
but has been developed mainly using spin-coefficients.
In order to make a comparison with such a description,
the foregoing results must be translated into spin-coefficients.
The notation of Penrose \& Rindler [9] is adopted;
equation numbers in triples indicate this reference.
Some minor generalisations are needed,
since the normalisation $\chi$ (2.5.46) cannot be set to 1 in this context.
The complex curvature may be defined by
$$K=\sigma\sigma'-\rho\rho'-\Psi_2+\Phi_{11}+\Pi\eqno(11)$$
as in (4.14.20).
Take a spin-basis adapted to a given spatial 2-surface $S$,
and propagate it away from $S$
such that the null normals are hypersurface-orthogonal,
$$0=\kappa=\kappa'=\rho-\bar\rho=\rho'-\bar\rho'\eqno(12)$$
as in (7.1.58).
The full set of `gauge conditions' corresponding to a double-null foliation
is given in [14], but is unnecessary for what follows.
Then the Hawking energy (1) takes the form
$$E=\B\I\up{K+\rho\rho'}\eqno(13)$$
with the area $A$ (2) as before.
Derivatives normal to $S$ may be expressed in terms of the weighted operators
$\thorn$ and $\thorn'$, (4.12.15).
Then $(K+\bar K)/\chi\bar\chi$ is the Gaussian curvature of $S$,
cf.\ (4.14.21),
and one has the Gauss-Bonnet theorem and its imaginary analogue,
$$\eqalignno
{&\I\up{K+\bar K}=4\pi(1-\gamma)&(14a)\cr
&\I\up{K-\bar K}=0&(14b)\cr}
$$
cf.\ (4.14.42--43).
Thus
$$\eqalignno
{&\thorn\I\up{K}=0&(15a)\cr
&\thorn'\I\up{K}=0.&(15b)\cr}
$$
Also, since $\thorn{*}1=-{*}2\rho$
and $\thorn'{*}1=-{*}2\rho'$ one has
$$\eqalignno
{&\thorn A=-2\I\rho&(16a)\cr
&\thorn'A=-2\I\rho'.&(16b)\cr}
$$
The focussing equations (6) correspond to
$$\eqalignno
{&\thorn\rho=\rho^2+\sigma\bar\sigma+\Phi_{00}&(17a)\cr
&\thorn\rho'=\eth\tau'+2\rho\rho'-\tau'\bar\tau'+K-\Phi_{11}-3\Pi&(17b)\cr
&\thorn'\rho=\eth'\tau+2\rho\rho'-\tau\bar\tau+K-\Phi_{11}-3\Pi&(17c)\cr
&\thorn'\rho'=\rho'^2+\sigma'\bar\sigma'+\Phi_{22}&(17d)\cr}
$$
using the field equations (4.12.32) and the definition of $K$ (11).
The dominant energy condition (5.2.9) implies
$$\eqalignno
{&\Phi_{00}\ge0&(18a)\cr
&\Phi_{11}+3\Pi\ge0&(18b)\cr
&\Phi_{22}\ge0.&(18c)\cr}
$$
Recall also (4.12.23),
$$(\thorn,\thorn',\eth,\eth')\chi=0.\eqno(19)$$
Then the variation (8) of the Hawking energy corresponds to
$$\eqalignno
{\thorn E&=\B
\left[\I\rho\up{K+\rho\rho'-\tau\bar\tau+\eth'\tau-\Phi_{11}-3\Pi}\right.\cr
&\qquad\qquad\qquad\left.+\I\rho'\up{\sigma\bar\sigma+\Phi_{00}}
-{1\over{A}}\I\rho\I\up{K+\rho\rho'}\right]&(20a)\cr
\thorn'E&=\B
\left[\I\rho'\up{K+\rho\rho'-\tau'\bar\tau'+\eth\tau'-\Phi_{11}-3\Pi}\right.\cr
&\qquad\qquad\qquad\left.+\I\rho\up{\sigma'\bar\sigma'+\Phi_{22}}
-{1\over{A}}\I\rho'\I\up{K+\rho\rho'}\right].&(20b)\cr}
$$
The definition of mean convex surface $S$ is equivalent to
$\O{\rho\rho'}<0$, and one may fix $\O\rho<0$,
so that $\thorn$ and $\thorn'$ differentiate
in the outgoing and ingoing directions respectively.
For a uniformly expanding foliation, $\O\rho$ and $\O{\rho'}$ are constant.
An equivalent statement of the monotonicity theorem is as follows,
with the positivity theorem being exactly as before.
\ms\ni
{\it Monotonicity theorem II.}
(i) For a uniformly expanding variation of a surface $S$,
the Hawking energy varies according to
$$\eqalignno
{&\thorn E=\B\I\up
{\rho'(\sigma\bar\sigma+\Phi_{00})-\rho(\tau\bar\tau+\Phi_{11}+3\Pi)}
&(21a)\cr
&\thorn'E=\B\I\up
{\rho(\sigma'\bar\sigma'+\Phi_{22})-\rho'(\tau'\bar\tau'+\Phi_{11}+3\Pi)}.
&(21b)\cr}
$$
(ii) For a mean convex surface $S$,
the dominant energy condition implies that
$E$ is non-decreasing in the outgoing null direction, $\thorn E\ge0$,
is non-increasing in the ingoing null direction, $\thorn'E\le0$,
and so is non-decreasing in any outgoing spatial direction,
and non-increasing in any ingoing spatial direction.
\ms\ni
{\it Proof.}
(i) Set $\O\rho$ and $\O{\rho'}$ to constants in (20),
and use the Gauss divergence theorem, $\I\eth'\tau=0$, (4.14.69).
(ii) Use the signs $\O\rho<0$, $\O{\rho'}>0$,
$\Phi_{00}\ge0$, $\Phi_{11}+3\Pi\ge0$ and $\Phi_{22}\ge0$ (18).
\ms
Note that rescaling both $\rho$ and $\rho'$ to constants
would be inconsistent with $\chi=1$ in general,
since $\rho\rho'/\chi\bar\chi$ is an invariant of the surface [14],
which is why $\chi$ has not been fixed.
A particular case of the formula (21) has been given by Eardley [15],
but assuming $\chi=1$,
for which (21a) and (21b) cannot both be generally valid.
\bs\ni
{\bf V. Asymptotics}
\ms\ni
In deriving the Bondi-Sachs energy loss as a limit of the Hawking energy loss,
a complication arises,
because the standard description uses a normalised spin-basis, $\chi=1$.
In this context one cannot choose a uniformly expanding foliation in general,
and so one must use the general energy-variation formula (20) rather than (21),
as follows.
A way to use the monotonicity theorem directly
is explained in the next section.

Following the description of \S9.8 of Penrose \& Rindler [9],
take a Bondi coordinate $u$ labelling a foliation of 2-surfaces on $\scri^+$,
and extend it inwards to label null hypersurfaces.
Choose an affine parameter $r$ on each null generator
which labels 2-surfaces such that $*1=\s r^2+O(1)$,
where $\s$ is the Hodge operator of the unit 2-sphere.
Take a spin-basis adapted to the 2-surfaces,
with $D=\partial/\partial r$ and $\chi=1$.
For an electrovac space-time,
$$\eqalignno
{&\Pi=0&(22a)\cr
&\Phi_{11}=\varphi_1\bar\varphi_1&(22b)\cr
&\Phi_{22}=\varphi_2\bar\varphi_2.&(22c)\cr}
$$
Assuming asymptotic simplicity (9.6.11),
the asymptotic expansions are given at the end of \S9.8 of Penrose \& Rindler.
The following are relevant:
$$\eqalignno
{&\rho=-r^{-1}+O(r^{-3})&(23a)\cr
&\rho'=\half r^{-1}+O(r^{-2})&(23b)\cr
&\sigma'=Nr^{-1}+O(r^{-2})&(23c)\cr
&\tau'=O(r^{-2})&(23d)\cr
&\varphi_1=O(r^{-2})&(23e)\cr
&\varphi_2=Fr^{-1}+O(r^{-2})&(23f)\cr
&K=\half r^{-2}+O(r^{-3}).&(23g)\cr}
$$
Then the area and Hawking energy behave as
$$\eqalignno
{&A=4\pi r^2+O(1)&(24a)\cr
&E=O(1).&(24b)\cr}$$
The Bondi-Sachs energy [1--2] may be defined as
$$\E(u)=\l E(u,r)\eqno(25)$$
cf.\ (9.9.59--61).
\ms\ni
{\it Bondi-Sachs energy-loss theorem I.}
For an electrovac, asymptotically simple space-time,
the Bondi-Sachs energy is non-increasing along a foliation of $\scri^+$,
the variation being given by
$$\thorn'\E=-{1\over{4\pi}}\II(N\bar N+F\bar F)\le0.\eqno(26)$$
\ms\ni
{\it Proof.}
Since $\thorn'$ commutes with the limit, $\thorn'\E=\l\thorn'E$.
The result follows from the variation formula (20b)
and the asymptotic expansions (23),
using the Gauss divergence theorem.
\ms
The complex function $N$
is called the {\it gravitational news}\/ function [1--2].
Similarly, $F$ may be called the {\it electromagnetic news}\/ function.
Despite appearances,
the news functions are quasi-local rather than local quantities,
since their definition involves the quasi-local quantity $r$,
as emphasised in \S9.10 of Penrose \& Rindler.
Similarly, there are no purely local news functions for the Hawking energy,
because the energy-variation formula (9) or (21)
is not the integral of a local energy flux,
but involves an overall factor in the area.
In a rougher sense, equation (21) shows that
the gravitational news for the Hawking energy
depends on $\sigma$, $\sigma'$, $\tau$ and $\tau'$,
of which only $\sigma'$ survives at $\scri^+$, yielding $N$.
It is interesting to note from (4.12.32) the direct relation
$\thorn'{*}\sigma'=*\Psi_4$ between the quasi-news $\sigma'$
and the gravitational radiation encoded in the Weyl spinor.

This method of deriving the Bondi-Sachs energy-loss formula
is somewhat more direct that that of Penrose \& Rindler,
due to the use of the Gauss-Bonnet theorem.
Penrose \& Rindler also give different definitions
of both the Bondi-Sachs energy and the news function,
which may be checked to be equivalent to $\E$ and $N$
using the asymptotic expansions.
The above definitions exhibit the relationship with the Hawking energy
more clearly,
and may therefore be regarded as more natural,
at least in the context of quasi-localisation.
\bs\ni
{\bf VI. Bondi-Sachs energy loss as a limit of Hawking energy loss}
\ms\ni
In this section,
an alternative derivation of the Bondi-Sachs energy-loss formula is sketched,
using the monotonicity theorem directly.
For a particular foliation of $\scri^+$ by conformal spatial 2-surfaces,
construct a double-null foliation in a neighbourhood.
Define the area radius
$$R=\sqrt{A/4\pi}.\eqno(27)$$
Then on purely geometrical grounds one has
$$\ll{*}R^{-2}=\s1\eqno(28)$$
and $\ll R^2\R=-\ll R^2\e^f\theta_+\theta_-=2$,
which may be strengthened to
$$\eqalignno
{&\R=2R^{-2}+O(R^{-3})&(29a)\cr
&\e^f=1+O(R^{-1})&(29b)\cr
&\theta_+=2R^{-1}+O(R^{-2})&(29c)\cr
&\theta_-=-R^{-1}+O(R^{-2})&(29d)\cr}
$$
where the rescaling freedom of the null normals has been used
to fix $R\theta_\pm$ at $\scri^+$.
Then $E=O(1)$, so that the Bondi-Sachs energy (25) is well defined.
Note that the cuts $S$ of $\scri^+$ are mean convex
in the obvious asymptotic sense,
$\O{R^2\theta_+\theta_-}<0$,
and that the foliation of $\scri^+$ is uniformly expanding,
since $\O{R\theta_+}=2$ and $\O{R\theta_-}=-1$ are constant.
In this sense, uniformly expanding foliations of mean convex surfaces
generalise cuts of $\scri^\pm$.
So the monotonicity theorem may be directly applied in this context,
provided the factors of $R$ cancel.
\ms\ni
{\it Bondi-Sachs energy-loss theorem II.}
(i) Assuming the asymptotic expansions (29),
and further that the limits
$$\eqalignno
{&\ll R^2\norm{\sigma_-}=8N^2&(30a)\cr
&\ll R^2\mod{\half\D f-\omega}=4B^2&(30b)\cr
&\ll R^2\phi_-=\Phi&(30c)\cr
&\ll R^2\rho=P&(30d)\cr}
$$
exist,
the variation of the Bondi-Sachs energy is well defined and given by
$$\L_-\E=-\II\left({1\over{4\pi}}(N^2+B^2)+\Phi+\half P\right).\eqno(31)$$
(ii) The dominant energy condition implies that
the Bondi-Sachs energy is non-increasing, $\L_-\E\le0$.
\ms\ni
{\it Proof.}
Apply the monotonicity theorem in the limit as $R\to\infty$,
substituting the asymptotic expansions (28--30).
\ms
Of course, it is preferable to derive the asymptotic expansions (30)
from simpler assumptions.
Using Penrose's definition of asymptotic simplicity, (9.6.11),
stricter versions of the expansions (28--30) are found,
and in particular, $B$ and $P$ vanish.
Consequently the dominant energy condition can be weakened to
the weak energy condition or the null energy condition
in the Bondi-Sachs energy-loss theorem.
The quantities $B$ and $P$ have been left undetermined in (30)
to allow for the possibility that they may be non-zero
under a different definition of asymptotic flatness,
for example, involving logarithmic terms in the asymptotic expansions.
\bs\ni
{\bf VII. Conclusion}
\ms\ni
A monotonicity property of the Hawking energy has been derived
which reduces at conformal infinity to the Bondi-Sachs energy loss.
In this sense, the Bondi-Sachs energy loss has been quasi-localised.
One may interpret this as meaning that
gravitational energy carries positive energy
in a wider regime than conformal infinity itself,
at least if
the Hawking energy can be regarded as measuring gravitational energy.
Since the assumptions of the monotonicity theorem hold
in a neighbourhood of $\scri^+$,
one may now measure gravitational energy
at a finite rather than infinite distance,
and be assured that it increases outwards and decreases inwards.
Also, the assumption of asymptotic flatness is now superfluous,
and the theorem may be used wherever mean convex surfaces can be found,
for instance, in a cosmological context or near a black hole.

The monotonicity theorem carries more information than the asymptotic version
in another sense:
the energy-variation formula (9) or (21) involves two terms,
of which one vanishes and one survives asymptotically,
but both of which have the same definite sign.
One of these signs might have been predicted from the conformal formula,
but the other is unexpected,
and could be taken to support the physical significance of the result.

Unfortunately, such interpretations are questionable
since the Hawking energy is non-zero for generic surfaces in flat space-time.
Thus one could find a mean convex surface in flat space-time
and calculate a positive flux of entirely fictitious gravitational radiation.
The Hawking energy can be corrected
by adding a term making it zero for any surface in flat space-time [7],
but the corresponding energy variation then contains extra terms
which do not have a fixed sign.\footnote\ddag
{The small-sphere behaviour of this Hamiltonian energy
has recently been calculated by Bergqvist [16],
confirming the lack of monotonicity in vacuum.
This energy and the Hawking energy,
and a 2-parameter family including both,
are the only suggested quasi-local energies
which are well defined for any embedded spatial 2-surface.}
One might interpret this as meaning that
gravitational radiation does not carry positive energy in general,
or does not make sense in general.

More conservatively, one might conclude that
the monotonicity and positivity theorems are physically relevant
only in situations
where the Hawking energy is a reasonable guide to gravitational energy,
for instance, for a surface close to spherical symmetry.
Indeed,
one might rescue the Hawking energy from its indefiniteness in flat space-time
by demanding that
it be used only for surfaces connectable by a uniformly expanding foliation
to either a point or a marginal surface,
or perhaps a suitably good cut [9] of $\scri$.
This suggests a concept of gravitational energy
which is not purely quasi-local,
but which requires a reference surface or point to provide a zero.
Although this is less generally applicable and conceptually less simple
than purely quasi-local energy,
it does give excellent results for the Hawking energy.

Finally, the potential usefulness of mean convex surfaces
should be emphasised.
As has been remarked, they are a type of converse of trapped surfaces,
characterising less exotic regions of space-time.
The positivity theorem uses mean convex surfaces to establish that
the Hawking energy satisfies the Penrose-Gibbons isoperimetric inequality
along a uniformly expanding foliation generated from a marginal surface.
If such a foliation can be extended to $\scri^+$,
then since the Hawking energy reduces to the Bondi-Sachs energy at $\scri^+$,
this would establish the isoperimetric inequality for the Bondi-Sachs energy,
and therefore also for the Arnowitt-Deser-Misner energy.
This would be a stronger property than mere positivity of the various energies,
and is regarded as a necessary condition for
the cosmic censorship hypothesis to hold [11--13].
A proof would follow from the convexity conjecture of Section II,
or something similar.
The conjectured inequalities are certainly true in spherical symmetry [17],
where the convexity conjecture holds.
On the other hand, the inequality which has been derived
is stronger than the conjectured versions in a different sense, namely that
it holds in general---not necessarily asymptotically flat---space-times.
\bs\ni
Acknowledgements.
It is a pleasure to thank
G\"oran Bergqvist, Tetsuya Shiromizu and James Vickers for discussions,
and the Japan Society for the Promotion of Science for financial support.
\np
\begingroup
\parindent=0pt\everypar={\global\hangindent=20pt\hangafter=1}\par
{\bf References}\ms
\ref{[1] Bondi H, van der Burg M G J \& Metzner A W K 1962}\PRS{A269}{21}
\ref{[2] Sachs R K 1962}\PRS{A270}{103}
\ref{[3] Penrose R 1982}\PRS{A381}{53}
\ref{[4] Hawking S W 1968}\JMP9{598}
\ref{[5] Hayward S A 1993}\CQG{10}{779}
\ref{[6] Hayward S A 1994}{General laws of black-hole dynamics, \PR}{D49}
{(in print)}
\ref{[7] Hayward S A 1994}\PR{D49}{831}
\refb{[8] Penrose R 1968 in}{Battelle Rencontres}
{ed: DeWitt C M \& Wheeler J A (New York: Benjamin)}
\refb{[9] Penrose R \& Rindler W 1986 \& 1988}
{Spinors and Space-Time Volumes 1 \& 2}{(Cambridge University Press)}
\ref{[10] Horowitz G T \& Schmidt B G 1982}\PRS{A381}{215}
\ref{[11] Penrose R 1973}\ANY{224}{125}
\refb{[12] Gibbons G W 1984 in}{Global Riemannian Geometry}
{ed: Willmore T J \& Hitchin N J (New York: Ellis Horwood Ltd)}
\ref{[13] Hayward S A, Shiromizu T \& Nakao K 1994}
{A cosmological constant limits the size of black holes, \PR}{D49}{(in print)}
\refb{[14] Hayward S A 1994}
{Spin-coefficient form of the new laws of black-hole dynamics}
{(in preparation)}
\refb{[15] Eardley D M 1979 in}{Sources of Gravitational Radiation}
{ed: Smarr L (Cambridge University Press)}
\refb{[16] Bergqvist G 1994}{Energy of small surfaces}
{(Karlstad University preprint)}
\refb{[17] Hayward S A 1994}{Gravitational energy in spherical symmetry}
{(in preparation)}
\endgroup
\bye